\input harvmac
\input epsf

\newcount\figno
\figno=0 
\def\fig#1#2#3{
\par\begingroup\parindent=0pt\leftskip=1cm\rightskip=1cm\parindent=0pt
\baselineskip=11pt
\global\advance\figno by 1
\midinsert
\epsfxsize=#3
\centerline{\epsfbox{#2}}
\vskip 12pt
{\bf Fig.\ \the\figno: } #1\par
\endinsert\endgroup\par
}
\def\figlabel#1{\xdef#1{\the\figno}}

\def\CM{{\cal M}}
\def\la{\langle}
\def\ra{\rangle}
\def\ph{\hat{\phi}}
\def\Gr{{\rm Gr}}
\def\kh{K\"{a}hler }
\def\bz{{\bar z}}

\def\bv{{\bar v}}

\def\p{\partial}

\font\cmss=cmss10
\font\cmsss=cmss10 at 7pt

\def\inbar{\vrule height1.5ex width.4pt depth0pt}
\def\IN{\relax{\rm I\kern-.18em N}}
\def\IP{\relax{\rm I\kern-.18em P}}
\def\IR{\relax{\rm I\kern-.18em R}}

\def\IZ{\relax\ifmmode\mathchoice
{\hbox{\cmss Z\kern-.4em Z}}{\hbox{\cmss Z\kern-.4em Z}}
{\lower.9pt\hbox{\cmsss Z\kern-.4em Z}}
{\lower1.2pt\hbox{\cmsss Z\kern-.4em Z}}\else{\cmss Z\kern-.4em
Z}\fi}
\def\IC{{\relax\,\hbox{$\inbar\kern-.3em{\rm C}$}}}

\lref\cds{
A.~Connes, M.~R.~Douglas and A.~Schwarz,
``Noncommutative geometry and matrix theory: Compactification on tori,''
JHEP {\bf 9802}, 003 (1998)
[arXiv:hep-th/9711162].
}

\lref\witsft{
E.~Witten,
``Noncommutative Geometry And String Field Theory,''
Nucl.\ Phys.\ B {\bf 268}, 253 (1986).
}

\lref\bergref{
S. Bergman, {\it The kernel function and conformal mapping},
American Mathematical Society, 1970.
}

\lref\sen{
A.~Sen,
``Descent relations among bosonic D-branes,''
Int.\ J.\ Mod.\ Phys.\ A {\bf 14}, 4061 (1999)
[arXiv:hep-th/9902105];
A.~Sen,
``Universality of the tachyon potential,''
JHEP {\bf 9912}, 027 (1999)
[arXiv:hep-th/9911116].
}

\lref\sw{
N.~Seiberg and E.~Witten,
``String theory and noncommutative geometry,''
JHEP {\bf 9909}, 032 (1999)
[arXiv:hep-th/9908142].
}

\lref\sch{
V.~Schomerus,
``D-branes and deformation quantization,''
JHEP {\bf 9906}, 030 (1999)
[arXiv:hep-th/9903205].
}

\lref\gms{
R.~Gopakumar, S.~Minwalla and A.~Strominger,
``Noncommutative solitons,''
JHEP {\bf 0005}, 020 (2000)
[arXiv:hep-th/0003160].
}

\lref\berezin{F.~A.~Berezin, ``Quantization,''
Izvestiya AN USSR, ser.math. 38 1116 (1974);
``General concept of quantization,''
Commun.\ Math.\ Phys.\  {\bf 40}, 153 (1975);
``Quantization in complex symmetric spaces,''
Izvestiya AN USSR, ser. math. 39 (1975) 363.
}

\lref\gqreview{
A.~Echeverria-Enriquez,
M.~C.~Munoz-Lecanda, N.~Roman-Roy and C.~Victoria-Monge,
``Mathematical foundations of geometric quantization,''
Extracta Math.\  {\bf 13}, 135 (1998)
[arXiv:math-ph/9904008].
}

\lref\perelomov{
A. Perelomov,
{\it Generalized Coherent States and Their Applications,}
Springer-Verlag, 1986.
}

\lref\math{
J. H. Rawnsley,
``Coherent states and K\"ahler manifolds,''
Quart.\ J.\ Math.\ Oxford (2) {\bf 28}, 403 (1977);
J.~Rawnsley, M.~Cahen and S.~Gutt,
``Quantization Of Kaehler Manifolds. I: Geometric 
Interpretation Of Berezin's Quantization,''
J.\ Geom.\ Phys.\  {\bf 7}, 45 (1990);
M.~Schlichenmaier,
``Deformation quantization of compact Kaehler manifolds by  
Berezin-Toeplitz quantization,''
[math.qa/9910137];
M.~Schlichenmaier,
``Berezin-Toeplitz quantization and Berezin 
transform,''
[math.qa/0009219].
}

\lref\kontsevich{
M.~Kontsevich,
``Deformation quantization of Poisson manifolds, I,''
arXiv:q-alg/ 9709040.
}

\lref\rt{
N.~Reshetikhin and L.~Takhtajan,
``Deformation quantization of K\"ahler manifolds,''
[math.QA/9907171].
}

\lref\hoppe{
M.~Bordemann, J.~Hoppe, P.~Schaller and M.~Schlichenmaier,
``Gl (Infinity) And Geometric Quantization,''
Commun.\ Math.\ Phys.\  {\bf 138}, 209 (1991).
}

\lref\mp{
M. I. Monastyrskii and A. M.  Perelomov,
``Coherent States and Bounded Homogeneous Domains,''
Sov.\ Phys.\ Dokl. \ {\bf 17}, 12 (1973).
}

\lref\kara{A. Karabagedov,
``Pseudo-\kh Quantization on Flag Manifolds,''
[dg-ga 970940].}

\lref\onofri{
E.~Onofri,
``Landau levels on a torus,''
Int.\ J.\ Theor.\ Phys.\  {\bf 40}, 537 (2001)
[arXiv:quant-ph/0007055].
}

\lref\komaba{
J.~A.~Harvey,
``Komaba lectures on noncommutative solitons and D-branes,''
arXiv: hep-th/0102076.
}

\lref\dnreview{
M.~R.~Douglas and N.~A.~Nekrasov,
``Noncommutative field theory,''
Rev.\ Mod.\ Phys.\  {\bf 73}, 977 (2002)
[arXiv:hep-th/0106048].
}

\lref\hklm{
J.~A.~Harvey, P.~Kraus, F.~Larsen and E.~J.~Martinec,
``D-branes and strings as non-commutative solitons,''
JHEP {\bf 0007}, 042 (2000)
[arXiv:hep-th/0005031];
K.~Dasgupta, S.~Mukhi and G.~Rajesh,
``Noncommutative tachyons,''
JHEP {\bf 0006}, 022 (2000)
[arXiv:hep-th/0005006].
}

\lref\ghs{
R.~Gopakumar, M.~Headrick and M.~Spradlin,
``On noncommutative multi-solitons,''
arXiv:hep-th/0103256.
}

\lref\ns{
N.~Nekrasov and A.~Schwarz,
``Instantons on noncommutative R**4 and (2,0) superconformal six
dimensional theory,''
Commun.\ Math.\ Phys.\  {\bf 198}, 689 (1998)
[arXiv:hep-th/9802068].
}

\lref\nikita{
N.~A.~Nekrasov,
``Trieste lectures on solitons in noncommutative gauge theories,''
arXiv:hep-th/0011095.
}

\lref\nakajima{
H. Nakajima,
{\it Lectures on Hilbert schemes of points
on surfaces}, American Mathematical Society University Lecture
Series, 1999;
K.~Furuuchi,
``Instantons on noncommutative R**4 and projection operators,''
Prog.\ Theor.\ Phys.\  {\bf 103}, 1043 (2000)
[arXiv:hep-th/9912047];
N.~A.~Nekrasov,
``Noncommutative instantons revisited,''
arXiv:hep-th/0010017.
}

\lref\sphere{
Y.~Hikida, M.~Nozaki and T.~Takayanagi,
``Tachyon condensation on fuzzy sphere and noncommutative solitons,''
Nucl.\ Phys.\ B {\bf 595}, 319 (2001)
[hep-th/0008023];
S.~Iso, Y.~Kimura, K.~Tanaka and K.~Wakatsuki,
``Noncommutative gauge theory on fuzzy sphere from matrix model,''
Nucl.\ Phys.\ B {\bf 604}, 121 (2001)
[arXiv:hep-th/0101102];
K.~Hashimoto and K.~Krasnov,
``D-brane solutions in non-commutative gauge theory on fuzzy sphere,''
Phys.\ Rev.\ D {\bf 64}, 046007 (2001)
[arXiv:hep-th/0101145];
Y.~Kimura,
``Noncommutative gauge theories on fuzzy sphere and fuzzy torus from
matrix model,''
Prog.\ Theor.\ Phys.\  {\bf 106}, 445 (2001)
[arXiv:hep-th/0103192];
C.~T.~Chan, C.~M.~Chen, F.~L.~Lin and H.~S.~Yang,
``CP(n) model on fuzzy sphere,''
Nucl.\ Phys.\ B {\bf 625}, 327 (2002)
[arXiv:hep-th/0105087].
}

\lref\kl{
S.~Klimek and A.~Lesniewski,
``Quantum Riemann Surfaces: II. The Discrete Series,''
Lett.\ Math.\ Phys.\  {\bf 24}, 139 (1992).
}

\lref\aps{
G.~Alexanian, A.~Pinzul and A.~Stern,
``Generalized Coherent State Approach to Star
Products and Applications to the Fuzzy Sphere,''
Nucl.\ Phys.\ B {\bf 600}, 531 (2001)
[arXiv:hep-th/0010187].
}

\lref\cdpsjt{
M.~Chaichian, A.~Demichev, P.~Presnajder, M.~M.~Sheikh-Jabbari and A.~Tureanu,
``Quantum theories on noncommutative spaces
with nontrivial topology:  Aharonov-Bohm and Casimir effects,''
Nucl.\ Phys.\ B {\bf 611}, 383 (2001)
[arXiv:hep-th/0101209].
}

\lref\torus{
E.~M.~Sahraoui and E.~H.~Saidi,
``Solitons on compact and noncompact spaces in large noncommutativity,''
Class.\ Quant.\ Grav.\  {\bf 18}, 3339 (2001)
[arXiv:hep-th/0012259];
I. Bars, H. Kajiura, Y. Matsuo and T. Takayanagi,
``Tachyon Condensation on Noncommutative Torus,''
Phys.\ Rev.\ {\bf D63} (2001) 086001,
[hep-th/0010101];
E. J. Martinec and G. Moore,
``Noncommutative Solitons on Orbifolds,''
[hep-th/0101199];
T.~Krajewski and M.~Schnabl,
``Exact solitons on noncommutative tori,''
JHEP {\bf 0108}, 002 (2001)
[arXiv:hep-th/0104090];
H.~Kajiura, Y.~Matsuo and T.~Takayanagi,
``Exact tachyon condensation on noncommutative torus,''
JHEP {\bf 0106}, 041 (2001)
[arXiv:hep-th/0104143].
}

\lref\gauge{
A. P. Polychronakos,
``Flux tube solutions in noncommutative gauge theories,''
Phys.\ Lett.\ {\bf B495} (2000) 407,
[hep-th/0007043];
D.~P.~Jatkar, G.~Mandal and S.~R.~Wadia,
``Nielsen-Olesen vortices in noncommutative Abelian Higgs model,''
JHEP {\bf 0009}, 018 (2000)
[hep-th/0007078];
D.~J.~Gross and N.~A.~Nekrasov,
``Dynamics of strings in noncommutative gauge theory,''
JHEP {\bf 0010}, 021 (2000)
[hep-th/0007204];
C.~Sochichiu,
``Noncommutative tachyonic solitons: Interaction with gauge field,''
JHEP {\bf 0008}, 026 (2000)
[hep-th/0007217];
R.~Gopakumar, S.~Minwalla and A.~Strominger,
``Symmetry restoration and tachyon condensation in open string theory,''
JHEP {\bf 0104}, 018 (2001)
[hep-th/0007226];
D. Bak,
``Exact Solutions of Multi-Vortices and False Vacuum Bubbles in
Noncommutative Abelian-Higgs Theories,''
Phys.\ Lett.\ {\bf B495} (2000) 251,
[hep-th/0008204];
M.~Aganagic, R.~Gopakumar, S.~Minwalla and A.~Strominger,
``Unstable solitons in noncommutative gauge theory,''
JHEP {\bf 0104}, 001 (2001)
[arXiv:hep-th/0009142];
J.~A.~Harvey, P.~Kraus and F.~Larsen,
``Exact noncommutative solitons,''
JHEP {\bf 0012}, 024 (2000)
[hep-th/0010060];
D.~J.~Gross and N.~A.~Nekrasov,
``Solitons in noncommutative gauge theory,''
JHEP {\bf 0103}, 044 (2001)
[arXiv:hep-th/0010090];
M.~Hamanaka and S.~Terashima,
``On exact noncommutative BPS solitons,''
JHEP {\bf 0103}, 034 (2001)
[hep-th/0010221].
}

\lref\scalar{D. Bak and K. Lee,
``Elongation of Moving Noncommutative Solitons,''
Phys.\ Lett.\ {\bf B495} (2000) 231,
[hep-th/0007107];
A. S. Gorsky, Y. M. Makeenko and K. G. Selivanov,
``On noncommutative vacua and noncommutative solitons,''
Phys.\ Lett.\ {\bf B492} (2000) 344,
[hep-th/0007247];
C. Zhou,
``Noncommutative scalar solitons at finite $\theta$,''
[hep-th/0007255];
U.~Lindstrom, M.~Rocek and R.~von Unge,
``Non-commutative soliton scattering,''
JHEP {\bf 0012}, 004 (2000)
[hep-th/0008108];
A. Solovyov,
``On Noncommutative Solitons,''
Mod.\ Phys.\ Lett.\ {\bf A15} (2000) 2205,
[hep-th/0008199];
B.~Durhuus, T.~Jonsson and R.~Nest,
``Noncommutative scalar solitons: Existence and nonexistence,''
Phys.\ Lett.\ B {\bf 500}, 320 (2001)
[hep-th/0011139];
M.~G.~Jackson,
``The stability of noncommutative scalar solitons,''
JHEP {\bf 0109}, 004 (2001)
[arXiv:hep-th/0103217].
}

\lref\splitsft{
L.~Rastelli, A.~Sen and B.~Zwiebach,
``Half strings, projectors,
and multiple D-branes in vacuum string field  theory,''
JHEP {\bf 0111}, 035 (2001)
[arXiv:hep-th/0105058];
D.~J.~Gross and W.~Taylor,
``Split string field theory. I,''
JHEP {\bf 0108}, 009 (2001)
[arXiv:hep-th/0105059].
}

\lref\rocek{
L.~Hadasz, U.~Lindstrom, M.~Rocek and R.~von Unge,
``Noncommutative multisolitons: Moduli spaces,
quantization, finite Theta  effects and stability,''
JHEP {\bf 0106}, 040 (2001)
[arXiv:hep-th/0104017].
}

\lref\av{
A.~Volovich,
``Discreteness in deSitter space and quantization of Kaehler manifolds,''
arXiv:hep-th/0101176.
}

\lref\hark{
J.~A.~Harvey and G.~W.~Moore,
``Noncommutative tachyons and K-theory,''
J.\ Math.\ Phys.\  {\bf 42}, 2765 (2001)
[arXiv:hep-th/0009030].
}

\lref\fs{
S.~Fredenhagen and V.~Schomerus,
``Branes on group manifolds, gluon condensates, and twisted K-theory,''
JHEP {\bf 0104}, 007 (2001)
[arXiv:hep-th/0012164].
}

\lref\ars{
A.~Y.~Alekseev, A.~Recknagel and V.~Schomerus,
``Open strings and noncommutative geometry of branes on group manifolds,''
Mod.\ Phys.\ Lett.\ A {\bf 16}, 325 (2001)
[arXiv:hep-th/0104054].
}

\lref\mumford{
D. Mumford,  {\it Tata Lectures on Theta I},
Birkh\"auser Boston, 1983.
}

\Title{\vbox{\baselineskip12pt\hbox{hep-th/0106180}
}}{Noncommutative solitons on \kh manifolds}

\centerline{Marcus Spradlin${}^\spadesuit$ and Anastasia
Volovich${}^\clubsuit$\foot{On
leave from the L. D. Landau Institute for Theoretical Physics,
Moscow, Russia.\smallskip
{\tt
${}^\spadesuit$spradlin@physics.harvard.edu},
{\tt
${}^\clubsuit$nastya@physics.harvard.edu}}}

\bigskip\centerline{Jefferson Physical Laboratory}
\centerline{Harvard University}
\centerline{Cambridge MA 02138}

\vskip .3in \centerline{\bf Abstract}
{
We construct a new class of scalar noncommutative multi-solitons on 
an arbitrary \kh manifold by using Berezin's
geometric approach to quantization
and its generalization to deformation quantization.
We analyze the stability condition which arises from
the leading $1/\hbar$ correction to the soliton energy and
for homogeneous \kh manifolds obtain that the 
stable solitons are given in terms of generalized coherent states.
We apply this  general formalism to a number of examples, 
which include the sphere, hyperbolic plane, torus and general
symmetric bounded domains.
As a general feature we notice that on homogeneous manifolds of
positive
curvature, solitons  tend to attract each other,
while if the curvature is negative they will repel each other.
Applications of these results are discussed.
}
\smallskip

\Date{June 2001}
\listtoc
\writetoc

\newsec{Introduction}

Starting from the celebrated paper
of Gopakumar, Minwalla and Strominger \gms, there
has been a lot of interest
in studying solitonic solutions of noncommutative
field theory.
Indeed,
noncommutative
geometry has arisen in at least three distinct but
closely related contexts in string theory.
Witten's open string field theory \witsft\ formulates
the interaction of bosonic open strings in the language of
noncommutative geometry.  Compactification of matrix theory
on the noncommutative torus \cds\ was argued to correspond to
supergravity with constant background three-form tensor field.
More generally, it has been realized \refs{\sch,\sw} that
noncommutative   
gauge theory arises as the worldvolume theory on D-branes in the 
presence of a constant background $B$ field in string theory.

Although Derrick's theorem forbids solitons in ordinary
2+1-dimensional scalar field theory, solitons in noncommutative
scalar field theory on the plane
were constructed in \gms\ (see also \refs{\nikita,
\komaba, \dnreview} for reviews).
It  was soon realized that noncommutative solitons represent
D-branes in string field  theory with a
background $B$ field turned on \refs{\hklm},
and this has allowed confirmation of some of
Sen's conjectures \sen\
regarding tachyon condensation in string field theory.
Other recent work on noncommutative solitons include
\refs{\scalar, \ghs, \rocek}
for scalar solitons and \gauge\ for solitons in 
noncommutative gauge theories.
Gauge theory solitons have been studied on the sphere in \sphere,
and on the torus in \torus.
Instanton solutions have been studied in \ns.

In this paper we develop a general approach to the
study of solitons in
noncommutative scalar field theory on any \kh manifold.
We define a noncommutative field theory
using the Berezin $\star$ product \berezin\
and its generalization to deformation 
quantization \refs{\kontsevich,\rt} and construct
static solutions of the theory.

The basic idea \gms\ behind the study of noncommutative solitons
is familiar by now.
One begins by exploiting an isomorphism between
the algebra of functions with the noncommutative $\star$ product and
the algebra of operators on some Hilbert space.
For example, the algebra of functions on the plane with the Moyal
$\star$ product can be represented as the algebra of 
operators on the Hilbert space ${\cal{H}} = L^2({\bf R})$.
In this sense the {\it coordinate} space on which the noncommutative
field theory is defined (i.e., the plane)
is identified as the {\it phase} space of
an auxiliary quantum mechanics (i.e., a particle on the line).
Noncommutative field theories defined in this way on compact
manifolds necessarily have a finite number of degrees of freedom,
since the corresponding Hilbert space is finite dimensional.
It is not yet clear whether it is possible to define  noncommutative
field theories with an infinite number of degrees of freedom
on curved compact spaces.

In a limit which corresponds roughly to infinite noncommutativity,
the potential term in the action dominates over the spatial part of
the kinetic term (henceforward referred to simply as the kinetic term,
since we will only consider static solutions).
Static
solutions to the approximate equation of motion $V'(\phi) = 0$ may
be constructed using projectors in the relevant operator
algebra.
Let us emphasize that we will also construct solitons based on
functions $\phi$ which
satisfy $\phi\star\phi=\phi$
in the formal sense of deformation quantization.  Since
there is no Hilbert space in this case, we will refer to these
as `projectors' rather than `projection operators.'\foot{Projectors
in a subalgebra of the
string field theory algebra
have recently been shown to play an
important role in the construction
of D-branes in split string field theory \splitsft.
}

One typically analyzes the stability of these
solutions by doing perturbation theory around infinite noncommutativity.
The space of rank $k$ projection operators is
isomorphic to the Grassmannian $\Gr(k,N)$ of complex $k$-planes
in $\IC^N$, where $N$ is the
dimension of the Hilbert space.
The leading  correction to the energy comes from the kinetic term,
which introduces an effective potential on $\Gr(k,N)$.
It was shown in \refs{\gms,\ghs,\rocek} that
on the plane,
any projection operator whose image is spanned by coherent states
minimizes the kinetic term, thus there is a moduli space for
solitons.\foot{Actually, the moduli space can be quite intricate
when two or more separated coherent state solitons are brought
together \ghs.  This is discussed more below.}

Note that in geometric quantization on a compact manifold, the Hilbert
space is always finite dimensional, so that the corresponding space
of projection operators $\Gr(k,N)$ is also compact.  Therefore the
kinetic term introduces a potential on $\Gr(k,N)$ which is bounded above
and below.
Depending on one's application, it may therefore not even be necessary
to consider minimizing that first perturbation to the energy.
By contrast on a noncompact manifold, the kinetic term is generally
unbounded, so that the perturbative expansion makes no sense
unless one is careful to analyze the contribution from the kinetic
term.

On a completely general manifold, the kinetic term has no symmetry and
thus one would not even expect a moduli space for a single soliton.
In this paper we show that on a homogeneous K\"ahler manifold (of
the type classified in subsection 5.4), there
is a moduli space for a single soliton represented by a projection
operator onto a generalized coherent state.

The organization of the paper is as follows.
In section 2 we review Berezin's approach to
the quantization of \kh manifolds and its
generalization to deformation quantization.
In section 3 we describe the noncommutative scalar field
theory that we will be studying and construct its
multi-soliton solutions as projectors in the corresponding algebra
of functions.
In section 4 we analyze the stability condition for these solitons,
which we are
able to solve for
homogeneous \kh manifolds, 
where
stable solitons are given in terms of generalized coherent states.
Sections 5 and 6 are devoted to examples.
In subsection 5.1 we connect our general formalism to the
very well-studied case of the plane.
The $\star$ product that we are using reduces on the plane
to the normal ordered operator product,
which is equivalent
to the more familiar Moyal product after a nonlocal field
redefinition.
In subsections 5.1--5.3 we study 
general symmetric bounded domains,
concentrating in great detail on the sphere and hyperbolic plane.
Among other things we show that 
that on a positive
curvature manifold solitons will tend to attract each other,
while  on a negative  curvature manifold they will repel each other
(although a discussion of why this might not
generalize to non-homogeneous manifolds is given in section 7).
We also show explicitly how the fuzzy versions of these manifolds
exactly map to Berezin $\star$ product.
Section 6 is devoted to the torus, where the breaking of
translational symmetry to $\IZ_N \times \IZ_N$ plays a crucial role.
We demonstrate the equivalence of the Berezin $\star$ product to the
fuzzy torus, and analyze a proposal for the definition of a kinetic term.
We conclude the paper in section 7 with a discussion of the
results and their possible applications.

\newsec{Quantization of \kh Manifolds}

In this section we briefly review Berezin's
approach
\berezin\ to the quantization of \kh manifolds using generalized
coherent states and its generalization
to deformation quantization \refs{\kontsevich, \rt}.\foot{Recently
this approach has also been used
in \av\ for quantization of the horizon in de Sitter space.
}

Let us recall the problem of quantization of a Poisson manifold
$\CM$.
Let ${\cal{A}}$
be the Lie algebra of smooth
functions on $\CM$ with the Poisson bracket
\eqn\pbracket{
\{ f, g \} = \omega^{ij} \p_i f \p_j g,
}
where $\omega$ is the Poisson structure on $\CM$.
Quantization is defined as
 a family $\CA_{\hbar}$ of
deformations of the algebra $\CA$,
where $\hbar$ is a parameter which takes values in some
subset of positive real numbers, such that $\CA_\hbar$
reduces to $\CA$ in the limit $\hbar \to 0$ in the following
sense:
\eqn\corr{\lim_{\hbar\to 0}f \star_\hbar g=fg,~~~~~
\lim_{\hbar\to 0}{1\over \hbar}(f \star_\hbar g-g \star_\hbar f)
=i\{f,g\},}
where $\star_\hbar$ is the multiplication operator in $\CA_\hbar$.
(Henceforward we will drop the $\hbar$ subscript on $\star$ for
notational simplicity.)

In geometric quantization
one associates to $\CM$ a family of Hilbert spaces
$\CH_\hbar$, with a correspondence between real-valued
functions in
$\CA_\hbar$ (i.e, observables) and hermitian operators $\CO_f$
on $\CH_h$ such that $\CO_{f \star g} = \CO_f \CO_g$.
The more general concept is deformation quantization 
\refs{\kontsevich,\rt}
where
the algebra
of functions is not necessarily representable as operators on any Hilbert
space.

In geometric quantization (for instance see \gqreview\ for a review),
the correspondence between functions $f$ and operators $\CO_f$
requires the introduction of a symplectic potential $\theta$
such that $\omega = d\theta$.
Of course $\theta$ will in general not be globally defined, so
the construction
depends on a local coordinate chart.
Since the action of a quantum operator on a wavefunction $|\psi\ra$
should be independent of the local chart, the Hilbert space
consists not of functions on $\CM$ but rather sections of a complex
line bundle $L$ over $\CM$.
Additionally one requires a ``polarization''
which roughly speaking is a splitting of the coordinates on the phase space
$\CM$ into those which are $q$'s (coordinates) and those which are $p$'s
(momenta).
The Hilbert space is then restricted to those wavefunctions which
depend only on the $q$'s.  On complex manifolds one takes
the complex structure as the polarization; then the Hilbert space $\CH$
is spanned by the holomorphic sections of $L$.

Berezin's approach to geometric quantization \berezin\ is based on the use
of generalized coherent states (see \perelomov\ 
for a review on coherent states)
and will be well-suited to
the study of noncommutative solitons.
His construction is also local, and therefore best suited for manifolds
for which there is an open dense subset which can be covered
with a single chart.
A global construction may be found in \math,
but will
not be necessary here.

Let $(M,\omega)$ be a \kh manifold, 
so that in local coordinates 
the metric and the \kh form are 
\eqn\mekah{
ds^2=g_{i\bar{k}}dz^idz^{\bar{ k}},  ~~~~~
\omega=g_{i{\bar k}}dz^i \wedge dz^{\bar k},
}
where $g_{i \bar{k}} = \p_i \p_{\bar{k}} K$ and $K$ is the K\"ahler
potential.

Berezin considered a complex line bundle $L$ over $\CM$ with 
a fiber metric $e^{-{1 \over \hbar}K(z,{\bar z})}$.\foot{We
will shortly see that for compact manifolds, a quantization condition
will relate the dimension of $\CH_\hbar$ to $\hbar$ and the volume
of $\CM$.}
The Hilbert space $\CH_\hbar$ therefore consists of holomorphic
sections of $L$ with the inner product
\eqn\scal{
\langle f| g \rangle
=c(\hbar)\int {\overline{f(z)}} g(z)
e^{-{1 \over \hbar}K(z,{\bar z})}
d\mu(z,{\bar z}),}
where $c(\hbar)$ is a normalization constant to be chosen shortly and
$d\mu$ is the measure 
\eqn\measure{
d\mu (z,\bz)=
\det|| g_{i {\bar k}}|| 
\prod {dz^k \wedge d \bz^k \over 2 \pi i}.
}

Let $\{f_k\}$ be a basis of holomorphic sections
orthonormal
with respect to \scal.
The basis may be finite or infinite, depending on whether
$\CM$ is compact.
The Bergman kernel \bergref\ is defined by
\eqn\bergman{B_{\hbar}(z,\bz)\equiv\sum_{k} f_k(z){\overline{f_k(z)}}.
}
It projects all measurable and square integrable functions
onto $\CH_{\hbar}$.
One can prove that it is independent of basis.
The holomorphic sections defined for $v \in \CM$ by
\eqn\coherent{
|v \rangle \equiv B_{\hbar}(\cdot,\bv)
}
are called generalized coherent states\foot{In
the global formulation, coherent states are parametrized not by
$\CM$ but rather by $L_0$, which is $L$ minus the image of the zero
section \hoppe.}
and
form an overcomplete system in $\CH_{\hbar}.$
We will use $\la v| = |v\ra^\dagger$ to denote
the antiholomorphic section $B_{\hbar}(v,\bz)$.
We choose the constant $c(\hbar)$ so that the resolution of
the identity reads
\eqn\resid{
{\bf 1} =  c(\hbar)
\int |v\ra \la v| e^{-{1 \over \hbar} K(v,\bar{v})} d\mu(v,\bv)
}
Note that for any holomorphic section $f(z)$ we have
\eqn\aaa{
\la v | f \ra = f(v).
}

For each bounded operator ${\hat O}$ on $\CH_{\hbar}$
we can define its symbol with respect to the
system  $|v \rangle$ as 
\eqn\symbdef{
O(z,\bv)={ \langle z |\hat O | v \rangle 
\over
\langle z| v \rangle}.}
Using \resid\ 
we can write a formula for the action of $\hat{O}$ on an arbitrary
section $f(z)$:
\eqn\symfun{\eqalign{
(\hat {O}f)(z) = \la z | \hat{O}|f \ra
 &= c(\hbar)
\int O(z,\bv)
\la z |v \ra \la v|f\ra 
e^{-{1 \over \hbar}
K(v,\bar{v})} d\mu(v,\bv)
\cr
&=
c(\hbar)
\int O(z,\bv)
B_{\hbar}(z,\bv) f(v) e^{-{1 \over \hbar} K(v,\bv)} d\mu(v,\bv).
}}

Functions in $\CA_{\hbar}$ are interpreted as symbols of operators
on $\CH_\hbar$.
The product of operators corresponds to the star product of
symbols.
Multiplication in $\CA_{\hbar}$ is defined as
\eqn\mult{({{O}}_1 \star {{O}}_2)(z,\bz)=c(\hbar)
\int {{O}}_1(z,\bv) {{O}}_2(v,\bz) 
{e_{\hbar} (z,\bv) e_{\hbar} (v,\bz) \over
e_{\hbar} (z,\bz)}
e^{{1 \over \hbar} \Phi(z,\bz|v,\bv)}
d\mu(v,{\bar v}),}
where  we have defined
\eqn\defehbar{e_{\hbar}(z,\bz)\equiv B_{\hbar}(z,\bz) e^{-{1 \over \hbar}
K(z,\bz)}}
and
$\Phi(z,{\bar z}|v,{\bar v})$ is the Calabi diastatic function
\eqn\yadr{\Phi(z,{\bar z}|v,{\bar v})=
K(z,\bv)+K(v,\bz)-K(z,\bz)-K(v,\bv),}
which is manifestly invariant under K\"ahler transformations and
therefore globally defined.

The trace of a symbol is defined by
\eqn\trace{
\Tr[O]=c(\hbar) \int O(z,\bz) e_{\hbar} (z,\bz) d\mu(z,\bz)
}
and is  cyclically invariant.
Substituting the identity operator, whose symbol is $1$, into \trace\
we find
\eqn\dimhil{
\dim \CH_{\hbar}=c(\hbar) \int e_{\hbar}(z,\bz)
d\mu(z,\bz).
}
Other properties of the $\star$ product are summarized in
the appendix.

Berezin was only able to
prove that his quantization procedure
satisfies the correspondence principle \corr\ under very restrictive
assumptions on the geometry of $\CM$ \berezin.  This quantization
works for the case when $\CM$ is
a homogeneous K\"ahler manifold and $L$ is a homogeneous bundle
(later it was generalized to generalized flag varieties \kara).
Under these assumptions $e_{\hbar} = 1$, and thus
most of the formulas simplify greatly.

Nevertheless, the formula \mult\ defines a $\star$ product on
any K\"ahler manifold in the formal sense of deformation
quantization \refs{\rt}.  That is, although it appears that the formulas
\symfun\ and
\mult\ only make sense for functions  which admit
an analytic continuation $f(z,\bv)$
to all of $\CM \times \CM$, in deformation
quantization these integrals are treated as formal power series
in $\hbar$.
Then the star product does not require
such an analytic continuation but depends only on the derivatives
(of all orders) of $f(z,\bar{z})$ evaluated on the diagonal.
Deformation quantization therefore works for all smooth functions,
and the star product reduces to Berezin's formula \mult\  for those functions
which do admit an analytic continuation.

\newsec{Noncommutative Solitons and Projectors}

We consider a noncommutative scalar field theory
on $\IR \times \CM$, where $\CM$ is an arbitrary
K\"ahler manifold of complex dimension $n$, with the action
\eqn\action{
S =
\int dt \int d\mu(z,\bar{z}) \left( {1 \over 2} \p_t \phi \star \p_t \phi
+ \phi \star \triangle \phi - m^2 V(\phi)_\star\right).
}
Here $\star$ is as defined in \mult\ and
$\triangle \equiv g^{i \bar{\jmath}} \p_i \p_{\bar{\jmath}}$
is the scalar Laplacian on $\CM$.
The subscript on the potential $V$ indicates that it is evaluated
with the star product.
We will assume that $V$ is bounded below, to ensure
that stable solutions exist.

Note that unlike the perhaps more familiar case of the Moyal product on the
plane, the star product cannot be omitted from the first term
in \action.
However in this paper we are interested in static
solutions of \action, so this fact will play no role and our task
will simply be to find minima of the energy functional
\eqn\funen{
E[\phi] \equiv \hbar^{n - 1}\int d\mu(z,\bz)\left(
-\phi \star \triangle \phi +  m^2 \hbar  V(\phi)_{\star}\right).
}
In writing this formula we have made use of the fact that by rescaling
the coordinates and size of $\CM$ one can formally eliminate $\hbar$
from the star product \mult.\foot{Instead, it will depend on the scale of 
curvature of $\CM$ (measured in
units of $\hbar$), or equivalently on the dimension 
of the Hilbert space, if $\CM$
is compact.}

One of the fascinating aspects of noncommutative scalar
field theory is that when the parameter $m^2 \hbar$
is sufficiently
large that the potential term dominates over the kinetic term in
\funen, then the extrema of $E[\phi]$ are insensitive to the exact
form of $V$.
To be concrete,
let us now assume that $V$ has a unique global minimum,
at $\phi = 0$, and a single local minimum
at $\phi = \lambda$.
Then
every solution $\phi_0$ to $V'(\phi_0)_\star = 0$
is of the form
$\phi_0 = \lambda \phi$ where $\phi$ satisfies
\eqn\proj{
(\phi \star \phi)(z,\bz) =
\phi(z,\bz).}
The problem of finding extrema of \funen\ therefore reduces in
the $m^2\hbar \to \infty$ limit to the problem of finding
projectors 
in the algebra $\CA_\hbar$ of functions with respect to the star
product \mult.

Projectors may be classified by their rank, and we will
begin with rank one.
A class of rank one projectors is given by
\eqn\resh{
\phi(z,\bz) = f(z) {C \over B(z,\bz)} \overline{f(z)},
}
where $f$ is an arbitrary holomorphic section, and
\eqn\cdef{
C^{-1} =
 c(\hbar) \int d\mu(v,\bv) f(v) \overline{f(v)} e^{-{1 \over \hbar}
K(v,\bv)}.
}
It is easily checked by direct substitution into
\mult\ that \resh\ formally satisfies $\proj$, and that its trace \trace\ is
indeed 1.

It is simple to generalize \resh\ to a rank $k$ projector
by choosing $k$ sections $f_i$ and defining the
matrix
\eqn\aaa{
h_{ij} = c(\hbar) \int d\mu(v, \bv) f_i(v) \overline{f_j(v)}
e^{-{1 \over \hbar} K(v,\bv)}.
}
Letting $h^{ij}$ denote the inverse of $h_{ij}$, the desired
rank $k$ projector
\eqn\mproj{
\phi(z,\bz) = \sum_{i,j} f_i(z) {h^{ij} \over B(z,\bz)}
\overline{f_j(z)}
}
satisfies \proj\ and has trace $k$.

Let us emphasize that the proceeding analysis has been
completely general, with all formulas holding in the formal
sense of deformation quantization,
allowing us to construct projectors on any
\kh manifold.
Of course for sufficiently complicated manifolds, such as
general Calabi-Yau
manifolds, it will not be possible to find explicit formulas for
the quantities appearing in \resh\ because not even the metrics on
these spaces are known.

In the special case when
the manifold admits a quantization where functions in
$\CA_\hbar$ on $\CM$ are represented as hermitian
operators on a Hilbert space, then $\resh$ is simply the symbol
of the operator $\ph = {|f\ra\la f| \over \la f|f\ra}$.
In this case it is clear that \resh\ is in fact the
most general
symbol of a
projection operator.
The symbol \mproj\ corresponds to the rank $k$
projection operator whose image is spanned by
the $|f_i\ra$.

\newsec{Stable Solitons}

When $m^2\hbar$ is large but still finite, we can
analyze the effect of the kinetic term in \funen\ by
doing perturbation theory in $1/(m^2\hbar)$.
The space of rank one projection operators in $\CA_\hbar$ is isomorphic
to the projective space
$\IP^N$, where $N = \dim \CH_\hbar$, and the kinetic
term
\eqn\kinetic{
E_1[\phi] \equiv - \lambda^2
\int d\mu(z,\bz) (\phi \star \triangle \phi)
}
defines a potential on $\IP^N$.
Note that the leading term in \funen\ is invariant under arbitrary
unitary transformations 
of the Hilbert space.
This U$(N)$ acts transitively on the space of projection operators
$\IP^N$.  The kinetic term \kinetic\ generically breaks this U$(N)$
symmetry, and in the following sections we will discuss in several
examples which subgroup (if any) of U$(N)$ is preserved by the kinetic
term \kinetic.

To find the minima of $E_1$ on $\IP^N$ we can use the method
of Lagrange multipliers to enforce the condition $\phi \star \phi = \phi$.
This yields the condition for the extrema
\eqn\uravvv{
\phi \star \triangle \phi=\triangle \phi \star \phi.
}
For rank one projection operators this is equivalent to
\eqn\urraaek{
(1-\phi) \star \triangle \phi \star \phi=0,
}
or
\eqn\urratwo{
\triangle \phi \star \phi = \alpha \phi
}
for some real number $\alpha$.  The
minima of $E_1$ are given by those $\phi$ which satisfy
\urratwo\ with the largest $\alpha$.

It seems difficult to study \urraaek\ and \urratwo\ for
general manifolds $\CM$,
but
for homogeneous K\"ahler manifolds, as discussed in the
previous section, an enormous simplification occurs because
$B(z,\bz) = e^{
K(z,\bz)}$.  In this case
we can take
a general soliton of the form \resh\ and
find
\eqn\lapphi{
\triangle \phi
= - n \phi +
 g^{i \bar{\jmath}}
(\nabla_i f) e^{-K} (\nabla_{\bar{\jmath}} \bar{f})
}
where $\nabla_i \equiv \p_i -
\p_i K$
is the covariant derivative
acting
on sections.
If we now substitute this into \urraaek\
we find that on homogeneous manifolds the
kinetic term \kinetic\ is minimized by rank
one projection operators of the form $|f\ra\la f|$ where
$f(z) = e^{K(z, \bv)}$ for some $\bv$, i.e.\ $|f\ra$
is a coherent state \refs{\perelomov,\mp}.
The associated projection operator \resh\ has symbol
\eqn\symcoh{
\phi_{v,\bv}(z,\bz) = e^{\Phi(z,\bz;v,\bv)},
}
where $\Phi$ is the Calabi function as defined in \yadr,
and can be thought of as a single soliton localized around
$z = v$.
Indeed on homogeneous manifolds the kinetic energy
\kinetic\ for a rank one projection operator $|f\ra\la f|$
is proportional  to the dispersion $(\Delta C)^2$
of the quadratic
Casimir $C$ of the group action in the state $|f\ra$,
and it is known \perelomov\ that the dispersion
is minimized by coherent states.
This will be discussed
more in section 5.

Let us now consider $k$-soliton configurations
which correspond to rank $k$ projection operators.
The space of such operators is the Grassmannian
$\Gr(k,N)$, and the kinetic energy \kinetic\ defines
a potential on $\Gr(k,N)$.
If we choose $k$ points $(v_i,\bv_i)$ on $\CM$, then
the projection operator whose image is spanned by the
$k$ associated coherent states $|v_i\ra$ has symbol
\eqn\mulaa{
\phi_{\{v_i, \bv_i\}}(z,\bz) = \sum_{ij} h_{ji} h^{ij}
e^{\Phi(z,\bz;v_j,\bv_i)}, ~~~~~ h_{ij} =  e^{K(v_i, \bv_j)}.
}
It was shown in \ghs\ that on the plane, the kinetic
energy is minimized in the space 
of rank $k$ projection operators precisely by those of the
form \mulaa\ whose image is spanned by coherent
states.\foot{See section 7 for a discussion of the subtleties
which arise
when two solitons come together.}
Thus there is a submanifold of the Grassmannian ($\Gr(k,
\infty)$ in that case)
which is an approximate moduli space (i.e., the energy
is minimized along that submanifold, but only at first
order in $1/(m^2 \hbar)$).

Note that throughout this paper, when we speak of
a projection operator associated to
$k$ coherent states, we mean $k$ separated coherent states.
We anticipate that the expression \mulaa\ has a perfectly
smooth limit when any two of the points are brought together.
This has been studied in detail on the plane \ghs, and the story
should be similar for any smooth manifold.

In the following sections we will by way of a number
of examples see that on general manifolds
the kinetic term is not constant on
any nontrivial submanifolds of $\Gr(k,N)$, so that
multi-soliton configurations will feel an effective
force which can either bring them together or push
them apart.

\newsec{Solitons on Homogeneous \kh Manifolds}

In this section we apply the analysis of the preceeding sections
to a number of homogeneous K\"ahler manifolds where explicit formulas
for multi-soliton solutions are easily given.
The relationship between Berezin's quantization and
`fuzzy' versions of various manifolds is also exploited.

\subsec{Plane}

Let us begin with
the very well studied example of the plane with  \kh potential 
\eqn\plkp{K(z,\bz)=z \bz.}
The Hilbert space is spanned by holomorphic functions of $z$
with inner product \scal
\eqn\aaa{
\langle f|g \ra = {1 \over
2 \pi} \int dz d\bz \ \overline{f(z)} g(z) e^{-z \bz}.
}
The creation and annihilation operators $a^\dagger$ and $a$ act
by differentiation and multiplication
\eqn\aaa{
(a f)(z) = {\p f \over \p z}, ~~~~~ (a^\dagger f)(z) = z f(z).
}

The $\star$ product \mult\ 
reduces to the well-known $\star$ product on the plane
which corresponds to Wick ordering of the operators:
\eqn\proizv{
({{O}}_1 \star {{O}}_2)(z,\bz)=
e^{\p_v \p_{\bv}}{{O}}_1(z,\bv) {{O}}_2(v,\bz)|_{v=z,\bv=\bz}.
}
There are five different ordering of operators on the plane:
$pq$, $qp$, Wick (or normal), anti-Wick, and Weyl ordering.
The symbols of operators in different orderings are related
by nonlocal field redefinitions.
For example, the
product of Weyl ordered symbols is the familiar Moyal product
\eqn\moyal{
({{O}}_1 \star {{O}}_2)(q,p)=
e^{{i  \over 2}
(\p_{q_1} \p_{p_2}-\p_{q_2} \p_{p_1})}{{O}}_1(q_1,p_1) {{O}}_2(q_2,p_2)
|_{q_1=q_2=q;p_1=p_2=p}.
}
The symbols ${{O}}_N(z,\bz)$
and ${{O}}_W(z,\bz)$ of a given operator with respect
to normal ordering and Weyl ordering respectively are related by
\eqn\svsym{
{{O}}_{W} (q,p)= 2 \int {{O}}_N (v,\bv)
e^{-2 (\bz -\bv)(z-v)} dv d\bv,
}
where
\eqn\svzip{
z={1 \over \sqrt{2}}(q+ip),~~~~~
\bz={1 \over \sqrt{2}}(q-ip).
}

The overcomplete basis of functions $|v\ra$
defined by \coherent\ 
are the usual coherent states $|v\ra = e^{\bv a^\dagger}|0\ra$.
Therefore a single soliton at a position $(v,\bv)$ corresponds
to the operator $\ph = {|v\ra\la v| \over \la v|v\ra}$,
whose symbol is
\eqn\planeone{
\phi_{v,\bv}(z,\bz) = e^{-(z-v)(\bz-\bv)}.
}
As already mentioned, it was shown in \ghs\ that
the kinetic term \kinetic\ is minimized
for rank $k$ projection operators by those
whose image is spanned by coherent states.
Since the kinetic term may be written in terms of operators
as
\eqn\planekin{
E_1[\ph] = \lambda^2 \Tr( [a,\ph][\ph,a^\dagger])
}
and is therefore independent of the ordering of symbols, this result
will still be true in our analysis.
That is, the operator $\ph$ which minimizes \planekin\ will
be the same as the one from \ghs, namely
\eqn\aaa{
P = \sum_{i,j} |v_i\ra h^{ij} \la v_j|, ~~~~~ h_{ij} = \la v_i
|v_j \ra = e^{\bv_i v_j},
}
but the corresponding normal-ordered symbol is
\eqn\aaa{
\phi_{\{v_i,\bv_i\}}(z,\bz)
= \sum_{ij}h_{ji} h^{ij}
e^{-(z-v_j)(\bz -\bv_i)}.
}
This corresponds to $k$ solitons localized at the positions
$(v_i,\bv_i)$.

\subsec{Sphere}

We can introduce complex coordinates $z,{\bar z}$ on a sphere of
radius $R$
by means of the stereographic projection
\eqn\coor{z=R \cot(\theta /2) e^{i\varphi},~~~~~
\bz=R \cot(\theta /2) e^{-i\varphi}.}
These coordinates cover the sphere except for the north pole.
In these coordinates the K\"ahler potential is
\eqn\kahsph{K=2 R^2 \ln (1 + {z \bz \over R^2}),
}
and the metric is
\eqn\metsph{
ds^2 = {4 \over (1+{z \bz \over R^2})^2} dz d\bz = 
R^2(d\theta^2+\sin^2\theta\,
d\varphi^2).}

Complex line bundles over the sphere are parametrized by a
single integer $n$ and denoted $\CO(n)$.  For $n \ge 0$, the bundle
$\CO(n)$ has no holomorphic sections and therefore would lead to
a trivial quantum mechanics.  For $n < 0$ the bundle has $N \equiv -n$
holomorphic sections, which may be written in the $z$-chart as
the functions $1,z,\ldots,z^{N-1}$.
Then the Hilbert space of geometric quantization
is $N$ dimensional.

The normalization constant $c$ appearing in the inner product
\scal\ is readily
found from the formula (A.2) to be $c = (1 + 2 R^2)/(4 \pi R^2)$.
Then \dimhil\ reads
\eqn\quantsph{
{\rm dim} \CH_\hbar = N = 2 R^2 + 1,
}
which is
the familiar quantization condition on the radius of a quantum
sphere.\foot{Of course \quantsph\ is usually phrased as a quantization
condition on $\hbar$.  Recall that we are working in units where $\hbar = 1$,
but upon restoring $\hbar$ by dimensional analysis, \quantsph\ reads
$\hbar = 2 R^2/(N-1)$.}

A single
soliton at a position $v$ on the sphere is described by the function
\eqn\sphss{
\phi_{v,\bv}(z,\bz)=
\left[{(1+{v \bz \over R^2})(1+{ z \bv \over R^2})\over 
(1+{v \bv \over R^2})(1+{z \bz \over R^2})}\right]^{N-1}.
}
By taking the limit $R \to \infty$, $N \to \infty$ while keeping
the relation \quantsph\ fixed we can go from the sphere to the
infinite plane.  In this limit the soliton \sphss\ becomes
\eqn\aaa{
\phi_{v,\bv}(z,\bz) = e^{-2 (z-v)(\bz - \bv)}.
}
(The factor of two difference between this and \planeone\ arises because
in our convention \kahsph\ the K\"ahler potential on the sphere
goes to $K = 2 z \bz$ in the flat space limit, as opposed to \plkp.)

Using \mulaa\ we can easily generalize \sphss\ to a function which
describes $k$ solitons at arbitrary positions $(v_i,\bv_i)$ on the sphere
(see Figure 1 for a graphical illustration).
Inserting such a function into the kinetic energy \kinetic, we find an
effective force which causes the two solitons to attract each other.
For example, for $N=3$ and $k=2$, with one soliton at the south pole
and the other at $z = \bz = r$, the attractive potential is
\eqn\atrr{
E_1(r) = {\lambda^2 \over R^2} \left[
2 + {2 \over (1 + {2 R^2 \over r^2})^2}
\right].
}

It is instructive to see how these results can be obtained
using the fuzzy sphere.
Indeed one can show explicitly \aps\
that the $\star$ 
product on the sphere defined by \mult\ corresponds to the ordinary
product of matrices of the fuzzy sphere, where the size
of the matrix is $N=1+ 2 R^2$.

Let us now review some basic facts about the fuzzy sphere.
The
Poisson bracket algebra of
the coordinate functions $x_a$ on the sphere
has the same structure as
the commutation relations for $SU(2)$ generators
\eqn\pgen{\{ x_a, x_b \}=R \epsilon_{abc} x_c.
}
Thus one can represent the Poisson bracket algebra
of the coordinates on the sphere by associating them to
matrices in the $N$ dimensional representation of SU$(2)$:
\eqn\otfm{x_a \to {2 R \over \sqrt{N^2-1}} \hat{J}_a,
}
where $[\hat{J}_a,\hat{J}_b] = i \epsilon_{abc} \hat{J}_c$ and
the normalization
is chosen so that $x_a x_a = R^2 {\bf 1}$.

The generators of the $SU(2)$ isometry in the $z$ coordinates
are
\eqn\generr{
J_3=\bz \p_{\bz}-z \p_z,
}
\eqn\generr{
J_+=-\p_z-\bz^2 \p_{\bz},~~~~J_-=\p_{\bz}+z^2 \p_z,
}
where $J_\pm \equiv J_1 \pm i J_2$.
If $O$ is the symbol of an operator $\hat{O}$, then $J_a O$
is the symbol of $[\hat{O},\hat{J}_a]$
The kinetic term \kinetic\ on the fuzzy sphere takes the form
\eqn\kinssph{
E_1[\ph]= {\lambda^2 \over R^2} \Tr ( 2 [\hat{J}_+,
\ph][\ph,\hat{J}_-]+[\hat{J}_3, \ph][\ph,\hat{J}_3]).
}
Written in this form it is manifest that the kinetic term 
breaks the U$(N)$ 
invariance of the potential term in \funen\ down to an
SU$(2)$ subgroup corresponding to overall rotations of the sphere.

For a rank one projection operator of the form $P = |\psi \ra
\la \psi|$ with $\la \psi | \psi \ra = 1$, \kinssph\ is simply
\eqn\dispersion{
E_1[|\psi \ra \la \psi|] = {2 \lambda^2 \over R^2}\, 
(\Delta \hat{J})^2,
~~~~~ (\Delta \hat{J})^2 \equiv
\la \psi| \hat{J}_a \hat{J}_a | \psi \ra
- \la \psi| \hat{J}_a |\psi\ra
\la \psi|\hat{J}_a|\psi \ra.
}
It is well known \perelomov\ that the dispersion $(\Delta
\hat{J})^2$
is minimized when $|\psi\ra$ is a highest (or lowest) weight state.
Thus the kinetic energy is minimized by projection operators
of the form $P= {|v \ra \la v | \over \la v| v \ra}$ which project
onto the coherent state
\eqn\sphvf{
|v \ra = e^{\bv \hat{J}_+} |j,\pm j\ra
}
built on a highest weight state
\eqn\aaa{
\hat{J}_3 |j,\pm j\ra = \pm j|j,\pm j\ra, ~~~~~
\hat{J}_\pm |j,\pm j \ra = 0.
}
The parameter $v$ comes from the SU$(2)$ 
invariance of \kinssph\ which allows us to move the soliton
to any point we like on the sphere.
The operator $P$ has symbol
\sphss.

In the fuzzy sphere formalism it
is easy to see that the \kinssph\ causes solitons to attract each other.
One soliton sitting on the north pole would be
described by
$P_1=|j,j\rangle \langle j,j|$,
and two solitons sitting on top of each other
by $P_2=|j,j \rangle \langle j,j|+|j,j{-}1 \rangle \langle j,j{-}1|.$
It is readily verified that $E_1(P_2) < 2 E_1(P_1)$, so that
there is a binding energy.
This is to be contrasted with the case of the plane \ghs, where a
Bogomolny bound ensured that $E_1(P_2) = 2 E_1(P_2)$.
In fact it is straightforward to derive the precise
potential \atrr\ from \kinssph\ using projection operators
on the fuzzy sphere.

\subsec{Lobachevsky Plane}

We have seen that on the plane, the kinetic term \kinetic\ 
allows for multi-solitons to sit at arbitrary positions (i.e., there is
a moduli space), while on the sphere, the kinetic term induces an
attractive potential between solitons.
In this subsection we consider the simplest homogeneous space with
negative curvature, namely the Lobachevsky plane.

The K\"ahler potential is
\eqn\aaa{
K(z,\bz) = - 2 R^2 \ln(1 - {z \bz \over R^2}),
}
where $R$ sets the radius of curvature and $(z,\bz)$ cover the disk
inside $z \bz < R^2$.
A single soliton at position $(v,\bv)$ is described by
\eqn\aaa{
\ph_{v,\bv}(z,\bz) =
\left[ { (1 - {v \bv \over R^2})(1 - {z \bz \over R^2}) 
\over (1 - {v \bz \over R^2})(1 - {z \bv \over R^2})} \right]^{2 R^2}.
}
Recall that $R$ is measured in units of $\hbar$, which we have set to
$1$.  Since
the space of holomorphic section is infinite dimensional,
there is no quantization condition, so $R$ and $\hbar$
are both free parameters of the model.

In this case one can easily check that the kinetic term \kinetic\ induces
a repulsive potential between two solitons.
As with the sphere, it turns out to be trivial to understand this result
by working with the fuzzy disk, which we now describe.

The group SU$(1,1)$ has a natural action on the complex plane.
Unlike the action of SU$(2)$, however, the SU$(1,1)$ action is not transitive
but instead foliates the plane into three orbits:  $X_+ = \{ z:
|z| < R\}$, $X_0 = \{ z: |z| = R\}$, and $X_- = \{ z : |z | > R\}$.
Also unlike SU$(2)$, which has a single type of unitary representation
(labelled by spin $j$), the group SU$(1,1)$ has three types of unitary
representations:  discrete, principal, and supplementary  (of course
all are infinite dimensional since SU$(1,1)$ is noncompact), which
are realized respectively on the space of functions on the orbit
of type $X_+$, $X_0$, and $X_-$.

So for the fuzzy disk we are interested in representations in the discrete
series.  These are labelled by a single number $k$ which takes
discrete values, $k=1,{3 \over 2},2,{5 \over 2},\ldots$.
Basis vectors $|k,m\ra$ are labeled by the eigenvalue of $K_0$,
\eqn\aaa{
K_0|k,m\ra = (k+m)|k,m\ra,
}
where $m$ is a non-negative integer.
The algebra is generated by operators $K_\pm$, $K_0$ satisfying
\eqn\comssoot{
[K_0,K_\pm]=\pm K_\pm,~~~~[K_-,K_+]=2 K_0.
}
The kinetic energy \kinetic\ on the fuzzy disk may be written as
\eqn\kindisk{
E_1[\ph]={\lambda^2 \over R^2} \Tr (-2[K_+, \ph][\ph,K_-]+[K_0, \ph][\ph,K_0])
}
and
is minimized among rank one projection operators by those which project
onto a coherent state
\eqn\dcohst{
|v\rangle =e^{\bv K_+}|k, 0 \rangle
}
built on a highest weight state.

One can check the repulsive force between two solitons by
comparing the energy of a single soliton at the origin given by
the projection operator $P_1 = |k,0\ra\la k,0|$ with the energy
of two solitons sitting on top of each other at the origin,
$P_2 = |k,0\ra \la k,0| + |k,1\ra \la k,1|$.  It is easily
verified that $E(P_2) > 2 E(P_1)$, so $P_2$ can lower its
energy by splitting apart and having the two solitons move far
away from each other.

\subsec{Symmetric Bounded Domains}

According to the Cartan classification
\berezin\
there are four types of  complex bounded symmetric domains
$M^I_{p,q}, M^{II}_p, M^{III}_p$ and $M^{IV}_n$,
and two exceptional domains.
The elements of domains are complex matrices $Z$
which satisfy the condition
\eqn\urdomain{
Z Z^{\dagger} < I,
}
where $Z$ is a
complex $p\times q$ matrix for $M^I_{p,q}$, symmetric $p\times p$ 
for $M^{II}_p$ and anti-symmetric $p\times p$ matrix for $M^{III}_p$.
Here $Z^{\dagger}$ is Hermitian conjugate and $I$ is the identity.
The fourth one $M^{IV}_n$ is given  by $n$-dimensional vectors.
Here we will consider domains of the first three types.
They are
$M^I_{p,q} = {SU(p,q) \over S(U(p)\times U(q))},$
$M^{II}_p = {Sp(p) \over U(p)}$ and
$M^{III}_p = {SO^*(2p) \over U(p)}.$
In particular $M^I_{1,1}=M^{II}_1
=S^2$ and $M^I_{1,q}$ is the complex projective space
$\IP^q$.

The \kh potential for the first three types is
\eqn\pote{K(Z,Z^\dagger)=\log \det(I-Z Z^\dagger)^{-\nu}}
where $\nu=p+q,p+1,p-1$ for $M^I_{p,q}, M^{II}_p, M^{III}_p$
respectively. 

We can apply the general formalism developed in sections
3 and 4 to construct solitons in these spaces.
The soliton at a position $v,\bv$ will be given by
\symcoh, where the \kh potential is given by
\pote.
Just like for the case of a sphere the solitons
on a positively curved coset will attract and 
those on a
negatively coset will repel each
other.\foot{We can also imagine taking direct products of coset spaces.
Then the Hilbert space will be a tensor product of two separate
Hilbert spaces, and multi-solitons in different components will
not talk to each other.}

This can be checked from the group theory
point of view using fuzzy cosets $G/H$.   As we discussed with SU$(1,1)$
above,
the only issue is to determine which representations of $G$ are
obtained in the fuzzy coset defined by Berezin's quantization.
The results are well-known \perelomov; for example for fuzzy
$\IP^2$ one uses $SU(3)$ representations of type $(m,0)$.

\newsec{Torus}

The torus is qualitatively different from the examples we have
discussed so far, because $e_{\hbar}(z,\bz) \ne 1$, so many
of the formulas remain complicated.
One manifestation of this is that the $U(1) \times U(1)$ isometry
of the torus will be `spontaneously' broken down to a $\IZ_N \times \IZ_N$
subgroup generated by discrete translations by $1/N$ times a lattice vector.
This symmetry breaking is a consequence of having a finite-dimensional
Hilbert space.\foot{Essentially, the problem is that
translational symmetry generators $[p_x,p_y] = i$
cannot exist in a finite-dimensional Hilbert space.  See \onofri\ for
a good discussion of this problem in the context of Landau levels on a
torus.}
The `fuzzy' torus studied here is therfore qualitatively
quite different from the `noncommutative' torus where
the Hilbert space is still infinite dimensional.
Noncommutative solitons on
the noncommutative torus have been studied in \refs{\ghs,\torus}.

\subsec{Holomorphic Sections and Fuzzy Torus}

For simplicity let us consider a rectangular torus
given by the quotient ${\bf T} \cong \IC/\Gamma$,
where $\Gamma = L(\IZ + \tau \IZ)$ is a lattice
on the plane and $\tau = i \tau_2$ with $\tau_2> 0$.
The constant $L$ sets the size of the torus (in units of $\hbar$).
In holomorphic gauge \onofri\ for the magnetic potential, the appropriate
K\"ahler potential is
\eqn\aaa{
K(z,\bz) = - {1 \over 4} (z - \bz)^2 = y^2.
}
We use the convention $z = x + i y$.
The quantization condition then reads
\eqn\aaa{
{\rm dim} \CH = N = {\tau_2 L^2 \over 2 \pi}.
}
In other words, the area of the torus is
$\tau_2 L^2  = 2 \pi N$, in agreement with
the Bohr-Sommerfeld quantization rule.
We choose the inner product on sections to be
\eqn\aaa{
\la f|g \ra = {1 \over N} \int_0^L dx \int_0^{\tau_2 L} dy
\  \overline{f(z)} g(z)\ e^{-y^2}.
}
This differs by an overall constant from the conventions of section 2,
but we will instead use the conventions of \hoppe\ in what follows.
A basis of orthonormal sections is given by
\eqn\aaa{
f_a(z) = {(2 N \tau_2)^{1/4} \over \sqrt{2 \pi}}
 \vartheta_{[a/N,0]}(z N/L, \tau N), ~~~~~ a=0,\ldots,N{-}1,
}
where
$\vartheta_{[a,b]}(z,\tau)$ is a Jacobi theta function
defined by
\eqn\oprtheta{
\vartheta_{[a,b]}(z,\tau)=\sum_{n \in \IZ}
e^{\pi i (n+a)^2 \tau +2 \pi i(n+a)(z+b)}.}
We will use $|a \ra$, $a=0,\ldots,N{-}1$
to denote the normalized basis vectors
corresponding to the holomorphic sections $f_a$.

Now consider the familiar SL$(2,\IZ)$ generators
on the torus with $\tau = i \tau_2$,
\eqn\aaa{
(S f)(z) = f(z + L), ~~~~~  (T f)(z) = e^{\pi i \tau/N^2 + 2 \pi i z/L}
f(z + \tau L/N).
}
These satisfy the commutation relations
\eqn\stcomm{
S T = e^{2 \pi i/N} T S.
}
One can easily check that the action of $S$ and $T$ in the basis
$|a\ra$ is given by \mumford
\eqn\stcomm{
 S | a\ra =  e^{2 \pi i a/N} |a \ra, ~~~~~
T | a \ra = |a + 1\ra.
}

The reader may recognize \stcomm\ as  the fundamental
algebraic
relation on the fuzzy torus, which is defined
as being the algebra generated by the matrices
\eqn\uvdef{
U = \left( \matrix{1 & & & \cr
 & \omega & & \cr
& & \ddots &\cr
& & & \omega^{N-1}}\right), ~~~~~
V = \left(\matrix{ & 1 & & \cr
& & 1 & \cr
& & & \ddots \cr
1 & & &}\right),
}
where $\omega = e^{2 \pi i/N}$, which satisfy
\eqn\uvcom{
VU = e^{2 \pi i/N} UV.
}
The geometric quantization of the torus presented here
is therefore precisely the fuzzy torus, and by comparing \uvdef\ to
\stcomm\ we immediately see that
\eqn\aaa{
S \leftrightarrow U, ~~~~~ T \leftrightarrow V^T
}
give the relation between the familiar clock and shift matrices $U$
and $V$ and the SL$(2,\IZ)$ generators acting on sections on the torus.
One can easily check that operator multiplication on the fuzzy torus
in the basis we are using maps to the star product \mult.

It is then an easy matter to write down noncommutative solitons
on the torus.  Let us start with a single soliton corresponding
to a projection operator onto a coherent state.  The Bergman
kernel is
\eqn\aaa{
B(z,\bz) =
{\sqrt{2 N \tau_2} \over 2 \pi}
\sum_{a=0}^{N-1} \left| \vartheta_{[a/N,0]} (z N/L, \tau N)
\right|^2.
}
The projection operator $\ph = {|v \ra \la v| \over \la v|v \ra}$
has symbol
\eqn\torusph{
\phi_{v,\bv}(z,\bz) = { B(z,\bv) B(v,\bz) \over B(z,\bz) B(v,\bv)}
}
This function represents a single soliton localized near
$(v,\bv)$.
One can check directly that the functional form of \torusph\ is invariant
only under discrete $\IZ_N \times \IZ_N$ translations which shift the
soliton by
$1/N$ times a lattice vector.  Figures 2 and 3 illustrate the lack
of translational symmetry graphically.
We will discover in the next subsection that localized
solitons of the form \torusph\ do not minimize the kinetic term and
are therefore unstable at first order in $1/(m^2 \hbar)$.

\subsec{Stable Solitons}

Because the translational symmetry is broken as discussed above,
even
defining the kinetic term \kinetic\ on the quantum torus is tricky.
The main problem is that \kinetic\ as written
is not well-defined, since
$\triangle \phi$ is not the symbol of any bounded operator even when
$\phi$ is, so that
$\phi \star \triangle \phi$ is undefined.
Another way to say this is that
$\triangle\phi$ does not admit an analytic
continuation to all of $\CM \times \CM$.  As discussed at
the end of section 2, this implies that
the formula \mult\ holds only in the
sense of deformation quantization, where $\hbar$ is treated as
a formal expansion parameter.

One might think that since we have shown how to recast the geometric
quantization of the torus into a simple algebra of $N \times N$ matrices,
it would be easy to overcome these difficulties.  However,
they persist even on the fuzzy torus, since 
one would like to define the kinetic term as on the plane by \planekin,
but the matrices $a$ and $a^\dagger$ cannot exist in a finite
dimensional algebra.
A candidate kinetic term for the fuzzy torus
has been presented in \refs{\dnreview,\cdpsjt}.
In our notation, it reads
\eqn\toruse{
E_1[\ph] = {\lambda^2 \over 2} \Tr\left[
(U \ph U^\dagger - \ph)^2 + (V \ph V^\dagger - \ph)^2
\right].
}
This kinetic term preserves the expected $\IZ_N \times \IZ_N$
subgroup of translational invariance,
\eqn\aaa{
E_1[ T_{ab} \ph T_{ab}^{-1} ] = E_1 [\ph], ~~~~~
T_{ab} \equiv U^a V^b.
}
Note that (up to a phase), $T_{ab}$ implements a discrete
translation by $(a/N,b/N)$.

It turns out that
the energy \toruse\ is minimized only by those projection operators
which either commute with $U$ or with $V$---that is, the image of
the projection operator must either be an eigenstate of $U$ or an
eigenstate of $V$.  The moduli space of minima of \torusph\ therefore
consists of $2N$ discrete points.  In position space these correspond
to a strip which is
localized around a $1/N$ lattice point in the $x$ (or $y$) direction
and extended along the $y$ (resp.\ $x$) direction.
It would be interesting to investigate the moduli space of multi-solitons
in more detail using our general approach.

\newsec{Discussion}

In this paper we 
constructed scalar noncommutative multi-solitons on 
an arbitrary \kh manifold by using Berezin's geometric approach to quantization
of \kh manifolds and its generalization to deformation quantization.
For homogeneous manifolds we analyzed stability conditions for these solitons
and showed that
stable solitons are given
in terms of  generalized coherent states.

We found that on homogeneous manifolds of positive curvature, coherent
state solitons tend to attract, while they repel each other on
homogeneous manifolds
of
negative curvature.  This is to be contrasted with the case of the
plane, where the leading correction \kinetic\ to the energy of the
solitons allows for a nontrivial moduli space.  It is tempting
to conjecture a general relation between curvature and the force
between solitons on general manifolds.  However, another important
ingredient in the story is whether the Hilbert space is finite or
infinite dimensional.

Indeed, it is easy to see quite generally that
projection operators whose image is spanned by generalized coherent
states at separated points cannot minimize the kinetic term \kinetic\ on
a compact
manifold.  This is because if $\ph$ minimizes the kinetic term in
the space of rank $k$ projectors ($\Gr(k,N)$),
then $1{-}\ph$ minimizes the kinetic
term in the space of rank $N{-}k$ projectors (which again is isomorphic
to $\Gr(k,N)$) and so it too must project onto coherent states.
Clearly $\ph$ and $1{-}\ph$ should be orthogonal, but generalized
coherent states are not orthogonal (in general).

It would therefore be interesting to investigate
the behavior of noncommutative solitons on   higher genus Riemann
surfaces $\CM_g$, which admit  negative curvature metrics but are compact
and therefore have a finite dimensional space of holomorphic
sections.
All such manifolds may be obtained as quotients of the hyperbolic
plane by some discrete group.
The quantization of these manifolds has been discussed
in \kl, where a basis of holomorphic sections was presented.

In \ghs\ it was shown that the moduli space of noncommutative
solitons on $\IC^d$ is the Hilbert scheme of $k$ points in $\IC^d$.
This means that the moduli space of solitons has a very interesting
structure when separated solitons are brought together---for $k>3$
and $d>2$ the moduli space is not even a manifold!
It would be interesting to see if there is a similarly rich structure
when generalized coherent states of the type used in this paper
come together and to explore the relation to Hilbert schemes \nakajima,
especially if there turns out to be a nontrivial moduli space for
noncommutative
solitons on any noncompact Ricci-flat surface.

There are many other directions for possible future investigation.
In would be  interesting to generalize these solitons 
to gauge theory and to explore their relation to D-branes
on \kh\ manifolds.
D-branes on group manifolds
have been constructed in WZW models
(see \ars\ for review),
and it would be interesting to explore the exact connection to those
constructions.
Also it would be interesting to compare the energy of
solitons to the D-brane tensions and to 
derive the properties of multi-solitons (attraction or repulsion)
that we obtained in section 5 from the D-brane 
point of view.

Finally,
noncommutative solitons have a very natural interpretation in
the
K-theory of $C^*$--algebras \refs{\hark, \fs} and it would be interesting to
understand
the K-theoretic interpretation of the solitons we constructed.

\bigskip

\centerline {\bf Acknowledgements}

\vskip .1in

We are grateful to M. Headrick, K. Hori, A. Maloney, S. Minwalla,
B. Pioline, M. Wijnholt and especially to R. Gopakumar for very 
useful discussions. We are also indebted to A. Strominger for
many useful discussions and comments on a preliminary
version of the paper.
This work is supported in part by DOE grant DE-FG02-91ER40654, and
A. V. is also supported by INTAS-OPEN-97-1312.

\appendix{A}{Properties of the $\star$ Product}

Here we will summarize the properties of the
$\star$ product, defined by \mult.
Plugging the operator $\hat{O}=1$
into \symfun\  
gives
\eqn\fourier{
f(z)=c(\hbar) \int f(v) e_{\hbar}(z,\bv)
e^{ {1 \over \hbar} K(z,\bv)
-{1 \over \hbar} K(v,\bv)} 
d\mu(v,\bv)= \langle z |f \rangle;
} 
from \mult\ since $1 \star 1=1$ 
we have finally
\eqn\intcal{
\int  {e_{\hbar} (z,\bv) e_{\hbar} (v,\bz) \over
e_{\hbar} (z,\bz)}
e^{{1 \over \hbar}\Phi(z,\bz|v,\bv)} d\mu(v,\bv) =c^{-1}(\hbar),
}  
and from $1 \star {{O}} = {{O}} \star 1= {{O}}$
\eqn\ravns{\eqalign{
{{O}}(z,\bz)&= c(\hbar) \int 
{e_{\hbar} (z,\bv) e_{\hbar} (v,\bz) \over
e_{\hbar} (z,\bz)}
{{O}}(z,\bv) e^{{1 \over \hbar} \Phi(z,\bz|v,\bv)}
d\mu(v,\bv)\cr
&=
c(\hbar) \int {{O}}(v,\bz) 
{e_{\hbar} (z,\bv) e_{\hbar} (v,\bz) \over
e_{\hbar} (z,\bz)}
e^{{1 \over \hbar} \Phi(z,\bz|v,\bv)}
d\mu(v,\bv).
}
}

\vskip 0.5in

\vfill
\eject

\appendix{B}{Figures}

{~}
\vskip .1in
{~}

\fig{
Two scalar solitons sitting at generic positions on a sphere
of radius $R = \sqrt{29/2}$ (in units of $\hbar$), which corresponds
by \quantsph\ 
to a Hilbert space of dimension $N =  30$.
As discussed in subsection 5.2, these solitons experience an attractive
force.
}{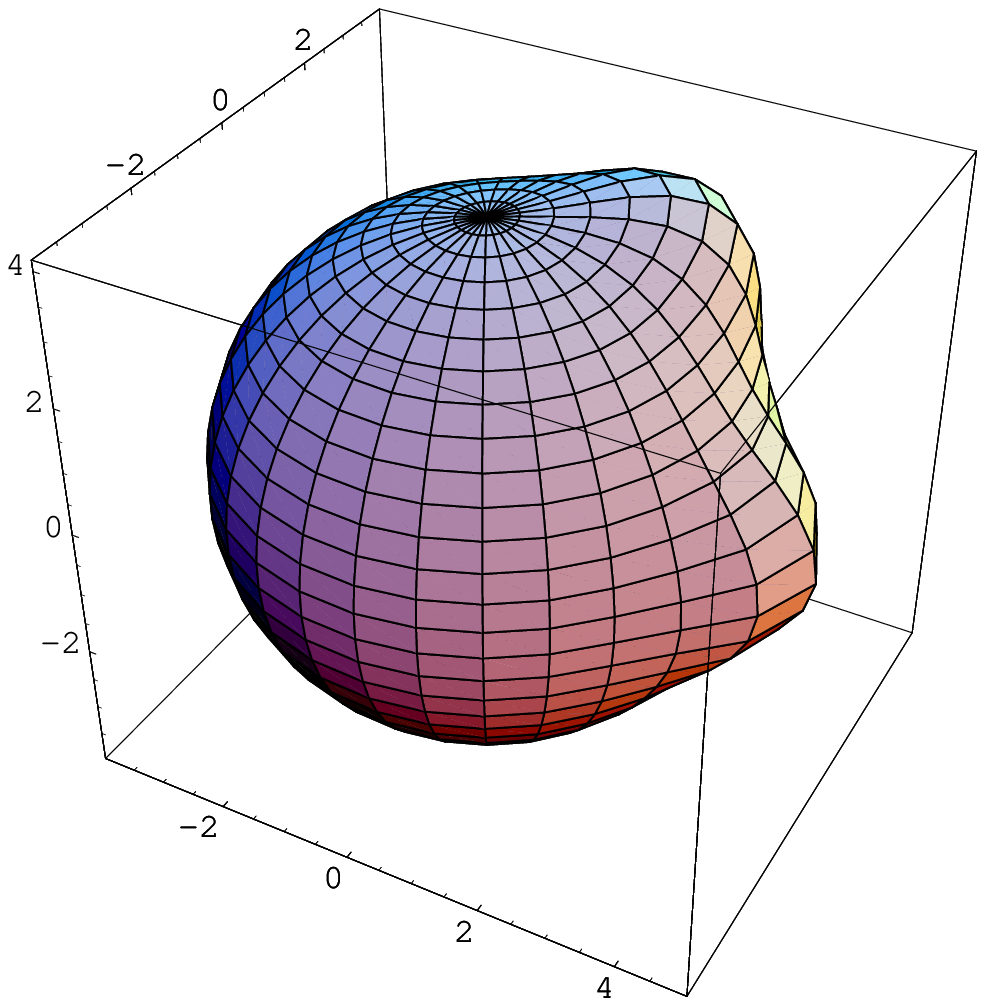}{1.5in}
\fig{
Four periodic copies of a single soliton on the square torus with $N = 4$
(thus $L = \sqrt{8 \pi}$ in units of $\hbar$),
given by \torusph\ with $v = 0$.
}{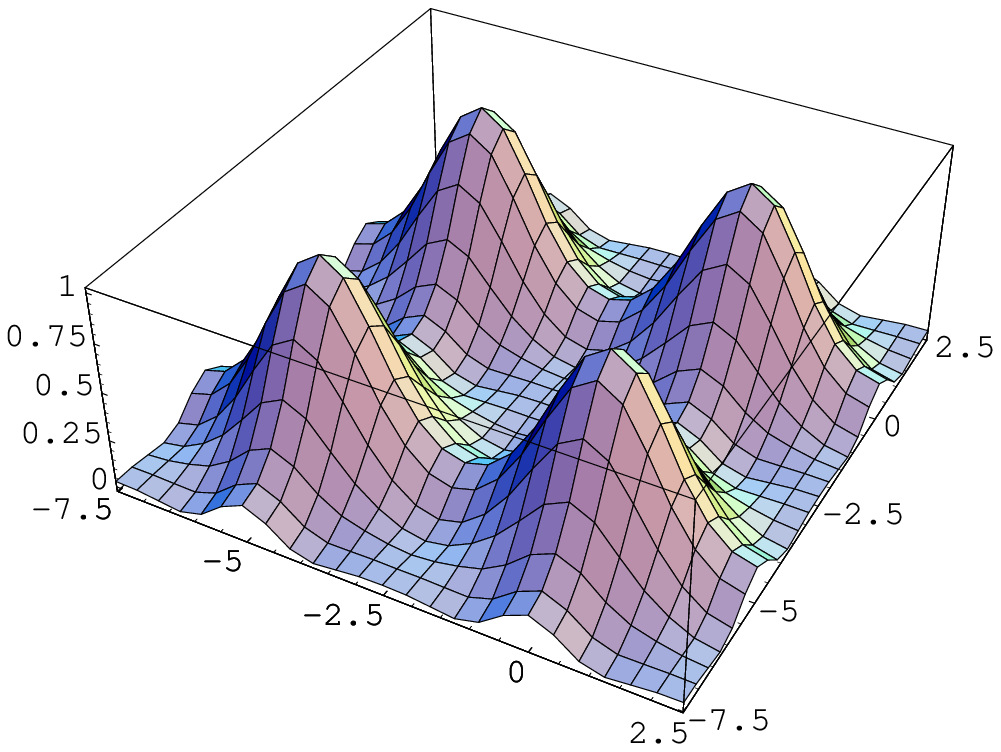}{1.7in}
\fig{
Same as in figure 2, but with $v = (1 +  i)L/2$.
The lack of translational symmetry is manifest.
}{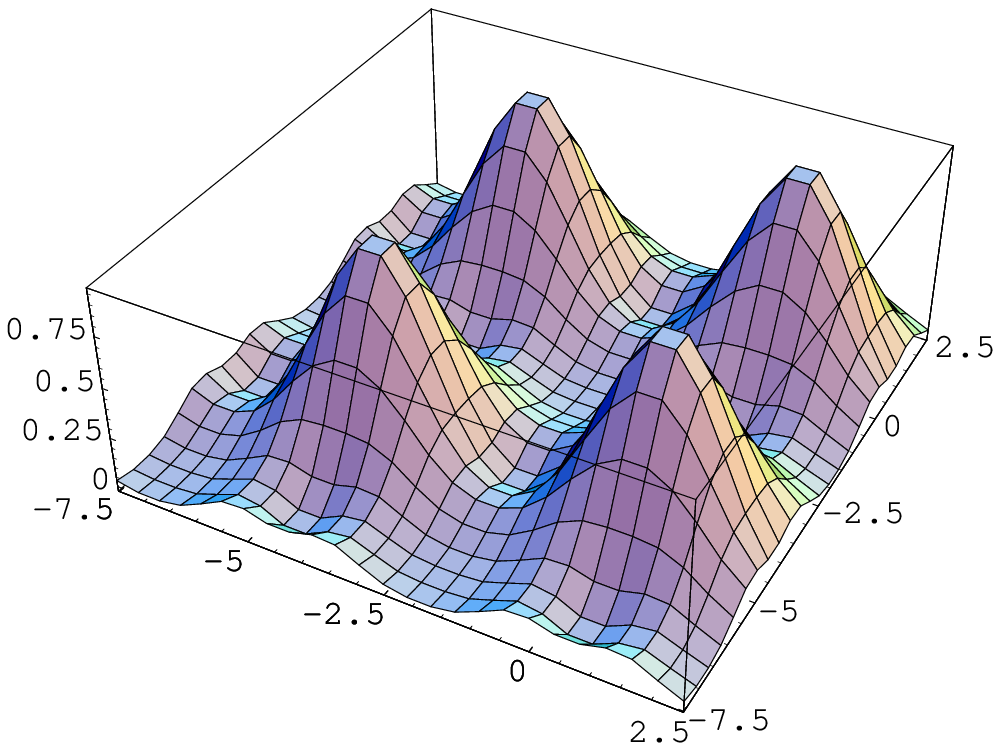}{1.7in}

\listrefs

\bye